\documentstyle[11pt,mrs2001,epsfig]{article}
\def\hMpc{h^{-1}{\rm Mpc}}
\def\etal{et al.}

\begin{document}

%
%\title{Tracing luminous and dark matter with the Sloan Digital Sky Survey}
\title{TRACING LUMINOUS AND DARK MATTER WITH THE SLOAN DIGITAL SKY SURVEY}

\author{J. LOVEDAY $^1$, for the SDSS collaboration}
\affil{$^1$Astronomy Centre, University of Sussex, Falmer, Brighton, BN1 9QJ,
England}

\begin{abstract}
I summarize the scientific goals and current status of the Sloan Digital
Sky Survey, briefly describe the Early Data Release, and discuss some recent
scientific results obtained from commissioning data which are apposite to 
the distribution of luminous and dark matter in the Universe.
\end{abstract}

\section{Introduction}

The Sloan Digital Sky Survey (SDSS) is the most ambitious effort to date
to map out the distribution of matter in the local Universe.
A technical summary of the survey is given by York et al.\ \cite{york2000}.
Very briefly, the survey hardware comprises the 2.5-metre survey telescope, 
a 0.5-metre photometric telescope (called the monitor telescope in
its previous incarnation),
a state-of-the-art imaging camera \cite{gunn98} that observes 
near-simultaneously
in five passbands $u'g'r'i'z'$ \cite{fuku96} and a pair of dual beam 
spectrographs, each capable of observing 320 fibre fed spectra.
The data are reduced by a series of automated pipelines and the
resulting data products stored in an object-oriented database.

The basic goal of the survey is to make a definitive map of the local Universe,
consisting of 5-colour imaging over $\pi$ sr to a depth $g' \approx 23$
and spectra of roughly one million galaxies and 100,000 QSOs.
In this contribution, I will discuss the current survey status,
the first public data release, and some science results obtained from
commissioning data.

\section{Survey Status}

First light with the imaging camera was obtained on 9 May 1998 and 
the first extra-galactic spectra were obtained in June 1999.
The survey was officially dedicated on 5 October 2000.
At the time of writing (July 2001), we have imaged 3135 square degrees
(24\% of the total survey area) and obtained spectra for 366 plug-plates,
yielding spectra for 158,000 galaxies, 20,000 QSOs and 27,000 stars,
including some repeated observations.
The spectrographs are performing extremely efficiently, with an overall
throughput of 20\% in the blue (3900--6000 \AA) and 25\% in the red
(6000--9100 \AA).
Automated spectral reduction pipelines classify these spectra and measure
redshifts.
In roughly 8\% of cases, the automated redshift measurement is in doubt and
the spectrum is flagged for human inspection.
About 1/8 of these (1\% overall) had their redshift manually corrected.
Based on manual inspection of all $\approx 23,000$ spectra from 39 plugplates,
this procedure correctly measures redshifts
for 99.7\% of galaxies, 98.0\% of quasars and 99.6\% of stars.

\section{The Early Data Release}

The first public release of SDSS data (hereafter EDR) took place on 
5 June 2001, and consists
of images covering 460 square degrees of sky, photometric parameters for
10 million objects and spectra for 55,000 objects.
The main access point to the data is through the website 
{\tt http://archive.stsci.edu/sdss/}.
The first paper to use this data \cite{gazt2001} was submitted to astro-ph
just 16 days after the release date!

\begin{figure}
\plottwo{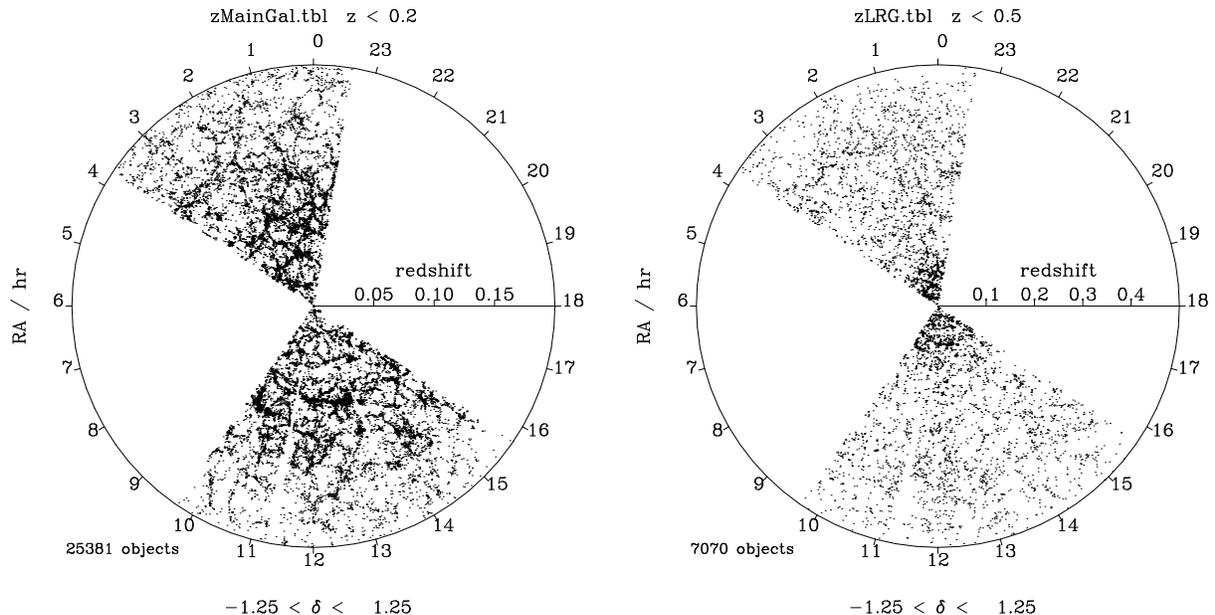}{loveday_fig1b.eps}
\caption{Distribution of EDR galaxies in RA and redshift around the equator
($|\delta| < 1.25\deg$).
The left plot shows 25,381 galaxies from
the main flux-limited galaxy sample within
a redshift $z=0.2$.
The right plot shows 7070 galaxies from the luminous red galaxy sample 
\cite{eisen2001} to $z = 0.5$.}
\label{fig:wedge}
\end{figure}

The distribution of equatorial galaxies in
the EDR is shown in RA-redshift wedge plots in Figure \ref{fig:wedge}.
The main galaxy sample is flux-limited ($r^* < 17.6$) and has a median redshift
$\bar{z} \approx 0.11$.
The luminous red galaxy (LRG) sample is designed to be volume-limited in the
redshift range $0.2 < z < 0.38$, and also includes galaxies to $z \le 0.5$.

\section{Early Science Results}

I highlight some early science results which provide important constraints
on the epoch of structure formation, the clustering of luminous matter
and the distribution of dark matter around galaxies.

\subsection{High-redshift QSOs}

The SDSS has broken the $z=6$ redshift barrier, with the the discovery of
a quasar at a redshift $z=6.28$, along with two new quasars at redshifts 
$z=5.82$ and $z=5.99$ \cite{fan2001}.
These objects were selected as $i$-dropouts: $i^* - z^* > 2.2$ and 
$z^* < 20.2$\footnote{The final SDSS photometry will be calibrated to
the system denoted $u'g'r'i'z'$.  The commissioning data is not yet
finally calibrated so the current magnitudes are indicated with asterisks:
$u^*g^*r^*i^*z^*$.}.
Contaminating L and T dwarfs were eliminated with followup near-IR 
photometry and confirming spectra were obtained with the ARC 3.5m telescope.
The SDSS has now observed a well-defined sample of four luminous quasars 
at redshift $z>5.8$.
The Eddington luminosities of these quasars are consistent with a 
central black hole of mass several times $10^9$ M$_\odot$,
and with host dark matter halos of mass $\sim 10^{13}$ M$_\odot$.
The existence of such mass concentrations at redshifts $z \approx 6$,
when the Universe was less than 1Gyr old, provides important constraints
on models of formation of massive black holes.
We expect to discover $\sim 27$  $z > 5.8$ quasars and one $z \approx 6.6$
quasar by the time the survey is complete.
Such observations will set strong constraints on cosmological models for
galaxy and quasar formation.

\subsection{Large scale structure}

\subsubsection{Angular clustering}

A series of papers \cite{conn2001, dod2001, scran2001, szalay2001, teg2001}
have studied the angular clustering of galaxies in SDSS commissioning data.
These papers are based on a single survey stripe (runs 752/756 observed in
March 1999) measuring $2.5 \times 90$ degrees and containing some 3 million
galaxies to $r^* = 22$.
Star-galaxy separation is performed using a Bayesian likelihood 
and approximately 30\% of the area is masked out due to poor seeing
\cite{scran2001}.
The angular correlation function, $w(\theta)$, is consistent with that 
measured from the APM Galaxy Survey \cite{mesl90} when scaled to the
same depth \cite{conn2001}.

\begin{figure}
\plottwo{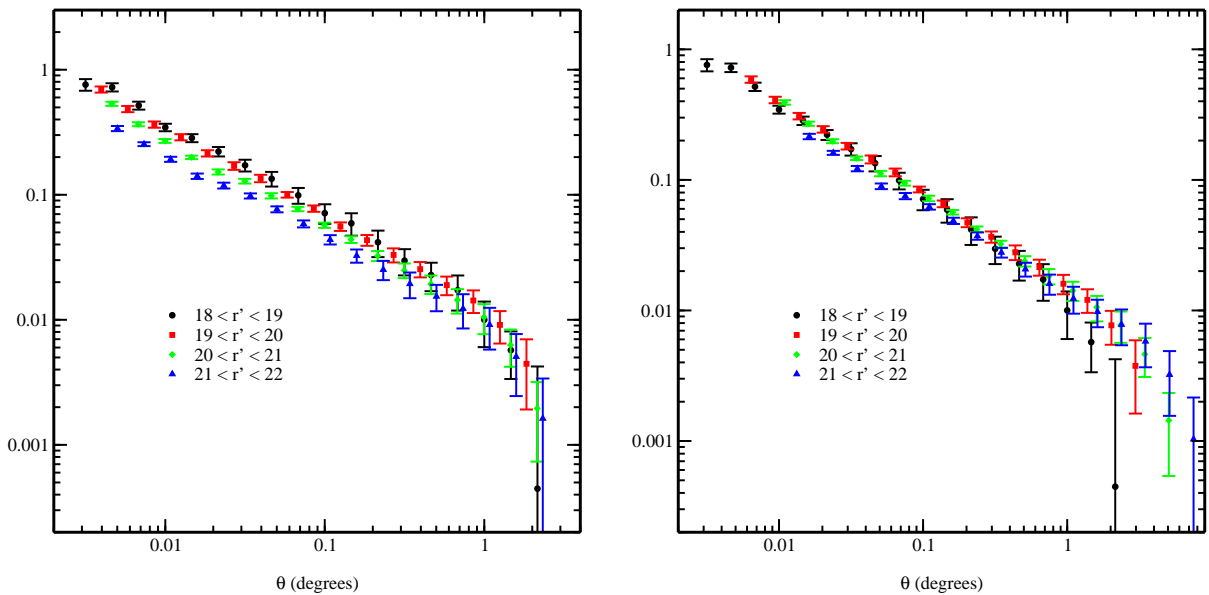}{loveday_fig2b.eps}
\caption{Scaling of the angular correlation function $w(\theta)$
measured in four
magnitude slices according to Limber's equation and assuming a
selection function assumed based on the CNOC2 survey \cite{lin99}.
The left plot assumes a $\Omega_m = 1$, $\Omega_\Lambda = 0$ cosmology,
the right plot a $\Omega_m = 0.3$, $\Omega_\Lambda = 0.7$ cosmology.
Assuming the latter cosmology improves the scaling of the faintest
($21 < r' < 22$) slice.
From~\cite{scran2001}.}
\label{fig:limber}
\end{figure}

An important test of the star-galaxy separation and of the photometric 
calibration is to check that $w(\theta)$ measured in magnitude slices
scales according to Limber's equation.
Figure \ref{fig:limber} shows that the scaling of $w(\theta)$ is well-described
by Limber's equation, particularly when an $\Omega_m = 0.3$, 
$\Omega_\Lambda = 0.7$ cosmology is assumed.
Further tests for possible sources of systematic errors in the SDSS data
are described in detail in \cite{scran2001} and the the angular clustering 
results are summarized in \cite{conn2001}.
Other papers describe estimates of the angular power spectrum \cite{teg2001},
inversion from $w(\theta)$ to the 3d power spectrum $P(k)$ \cite{dod2001} 
and direct estimation of power spectrum parameters \cite{szalay2001}.

\subsubsection{Spatial clustering}

\begin{figure}
\plottwo{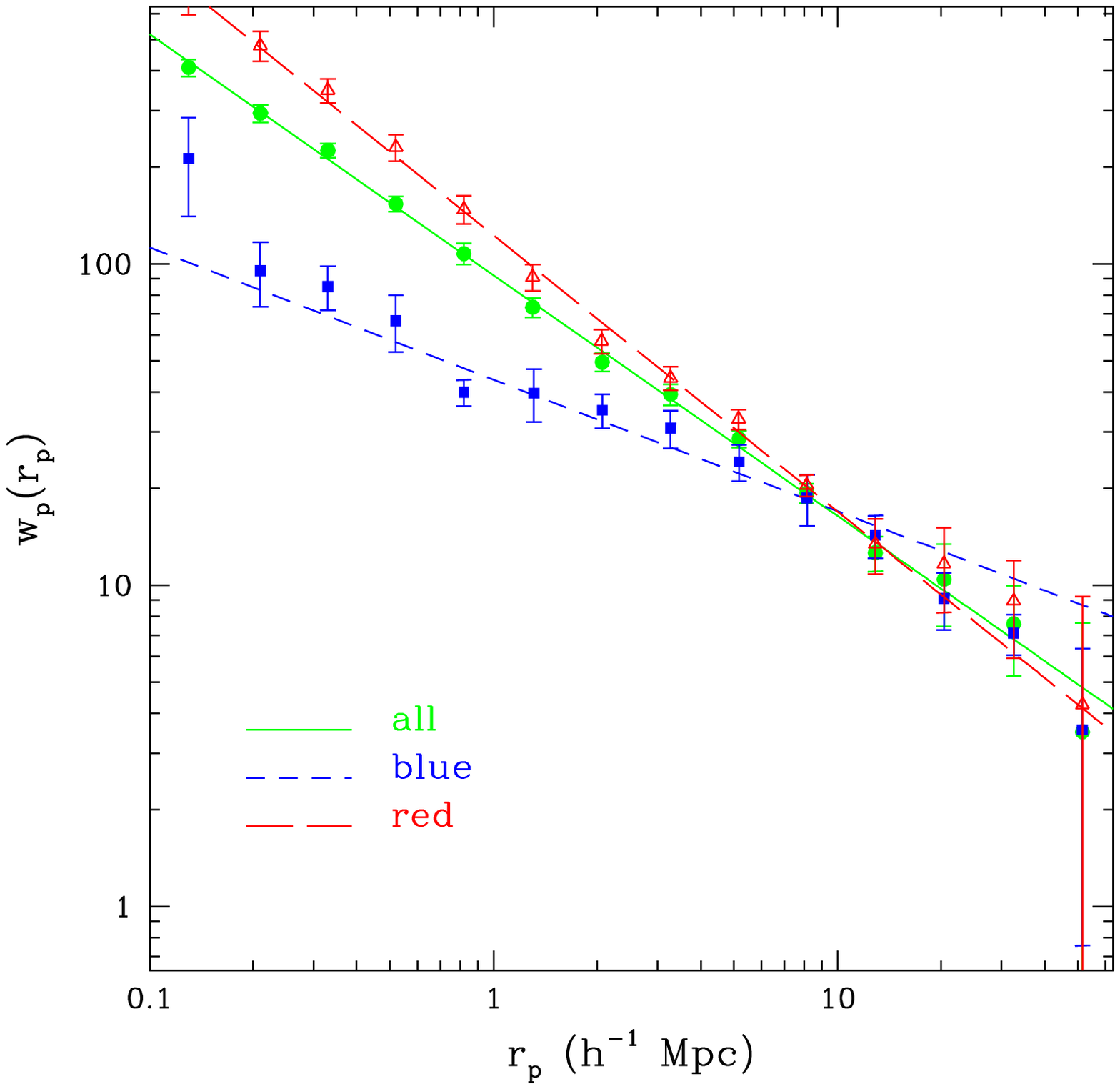}{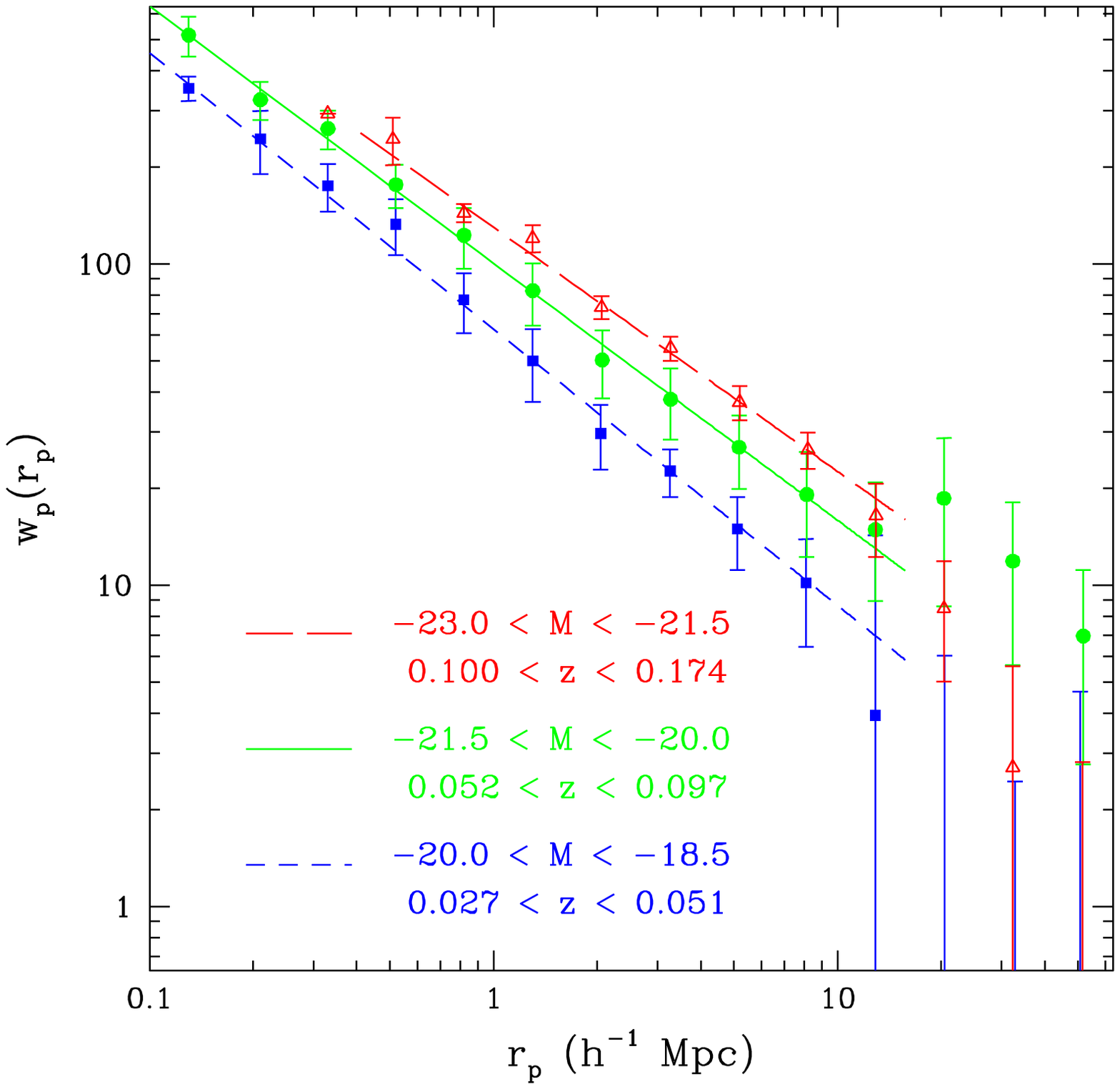}
\caption{Projected correlation functions $w_p(r_p)$ for redshift survey
galaxies subdivided by colour (left plot) and luminosity (right plot).
Note that the slope of $w_p(r_p)$ increases from blue to red colour,
but remains approximately constant with luminosity.
From~\cite{zehavi2001}.}
\label{fig:xi}
\end{figure}

A preliminary estimate of spatial clustering of galaxies has been made
using redshift information \cite{zehavi2001}.
This sample consists of 29,300 galaxies with $r^* < 17.6$ and within $\pm 1.5$
magnitudes of the characteristic magnitude $M_r^*$, distributed 
non-contiguously over 690 square degrees.
The real-space correlation function $\xi(r)$ is estimated by integrating
$\xi(r_p, \pi)$ over the line of sight separation $\pi$, where $r_p$ is 
the projected separation of two galaxies on the sky.
We find that $\xi(r)$ is well-fit over the range $0.1 < r < 30 \hMpc$
by a power law $\xi(r) = (r/r_0)^{-\gamma}$ with parameters given in
Table~\ref{tab:corr}.
The correlation length $r_0 = 6.14 \hMpc$ is slightly larger than the
$r_0 \approx 5.5 \hMpc$ found by earlier studies, presumably because dwarf
galaxies with $M > M^* + 1.5$ have been excluded from this analysis.
The pairwise velocity dispersion $\sigma_{12}$ is consistent with 600 km/s
for separations $r_p < 5 \hMpc$.

\begin{table}
\begin{center}
\caption{Power-law parameters for the real-space correlation function
$\xi(r) = (r/r_0)^{-\gamma}$.  Units for the correlation length $r_0$
are $\hMpc$.  From~\cite{zehavi2001}.}
\label{tab:corr}
\begin{math}
\begin{array}{lll}
\mbox{Sample} & \multicolumn{1}{c}{r_0} & \multicolumn{1}{c}{\gamma}\\
\mbox{All} & 6.14 \pm 0.18 & 1.75 \pm 0.03\\
\mbox{Red} & 6.78 \pm 0.23 & 1.86 \pm 0.03\\
\mbox{Blue} & 4.02 \pm 0.25 & 1.41 \pm 0.04\\
M^* - 1.5   & 7.42 \pm 0.33 & 1.76 \pm 0.04\\
M^*         & 6.28 \pm 0.77 & 1.80 \pm 0.09\\
M^* + 1.5   & 4.72 \pm 0.44 & 1.86 \pm 0.06
\end{array}
\end{math}
\end{center}
\vspace{-5mm}
\end{table}

Figure~\ref{fig:xi} shows the clustering properties for two subsamples of the
galaxy population selected by restframe $u-r$ colour at $(u^*-r^*)_0 = 1.8$,
corresponding roughly to bulge (red) and disk (blue) dominated galaxies.
The red galaxies exhibit a steeper power-law slope and longer correlation
length than the blue galaxies, as indicated by the power-law fit parameters
in Table~\ref{tab:corr}.
Also shown in Figure~\ref{fig:xi} are the correlation functions for three 
volume-limited
samples, centered on $M^* - 1.5$, $M^*$ and $M^* + 1.5$.
The power-law slopes for these samples are all consistent with $\gamma = 1.8$, 
although the correlation length $r_0$ decreases as 
expected from bright to faint luminosities.

\subsection{Galaxy-mass correlation function}

\begin{figure}
\parbox{95mm}{
\plotone{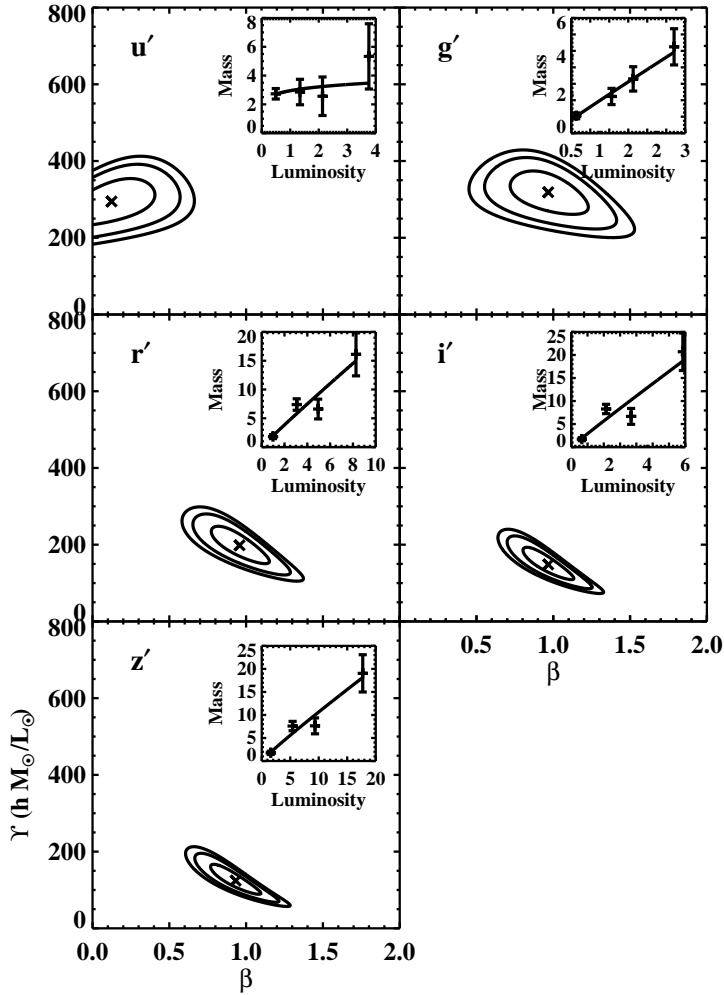}{95mm}
}
\hfill
\parbox{7cm}{
\caption{Mass-luminosity relation in the five SDSS bands estimated from 
weak lensing.
The inset in each panel plots estimated mass within $260 h^{-1}$kpc ($M_{260}$)
as a function of lens luminosity.
The contours show 1, 2 and 3 sigma confidence limits on the parameters
$\Upsilon$ and $\beta$ in the relation
$M_{260} = \Upsilon(L/10^{10}L_\odot)^\beta$.
Note that inferred mass has only very weak dependence on $u$-band luminosity,
but in the redder survey bands $griz$, the mass-luminosity relation appears 
to be linear.
From~\cite{mckay2001}.}
\label{fig:mckay}
}
\end{figure}

So far, I have summarized recent SDSS results concerning the distribution
of {\em luminous} matter in the Universe.
Direct constraints on the {\em dark} matter distribution may be obtained
from gravitational lensing.
McKay et al. \cite{mckay2001} have made weak lensing measurements of the
surface mass density contrast around foreground galaxies of known redshift.
Although the lensing signal is too weak to detect about any single lens,
by stacking together around 31,000 lens galaxies a clear lensing signal is
detected.
The galaxy-mass correlation function is well fit by a power-law of the form
$\Delta\Sigma_+ = 2.5 (r/\mbox{Mpc})^{-0.8} h$ M$_\odot$ pc$^{-2}$.
The strength of correlation is found to increase with the following 
properties of the lensing galaxy: late $\rightarrow$ early-type morphology,
luminosity in all bands apart from $u'$, and local density.
Figure~\ref{fig:mckay} shows the relationship between inferred mass within
a $260 h^{-1}$kpc radius and luminosity in each of the survey bands.

\section{Conclusions}

The Sloan Digital Sky Survey is now fully operational and is producing
high quality data at a prodigious rate.
We have imaged 3135 deg$^2$ of sky in five colours and have obtained
more than 200,000 spectra.
Much exciting science has already come out of just a small fraction of
the final dataset and we look forward to many more exciting discoveries in
the coming years.

\acknowledgements{The Sloan Digital Sky Survey (SDSS) is a joint
project of The University of Chicago, Fermilab, the Institute for
Advanced Study, the Japan Participation Group, The Johns Hopkins
University, the Max-Planck-Institute for Astronomy (MPIA), the
Max-Planck-Institute for Astrophysics (MPA), New Mexico State
University, Princeton University, the United States Naval Observatory,
and the University of Washington. Apache Point Observatory, site of
the SDSS telescopes, is operated by the Astrophysical Research
Consortium (ARC).

Funding for the project has been provided by the Alfred P. Sloan
Foundation, the SDSS member institutions, the National Aeronautics and
Space Administration, the National Science Foundation, the
U.S. Department of Energy, the Japanese Monbukagakusho, and the Max
Planck Society. The SDSS Web site is {\tt http://www.sdss.org/}.}

Many thanks to Laurence and Marie for organizing a most enjoyable 
and stimulating meeting.

\vfill
\end{document}